\documentstyle[12pt,aasms4,graphics]{article}

\begin{document}
\oddsidemargin=0mm

\title{Einstein's Biggest Blunder? High-Redshift Supernovae and the 
Accelerating Universe
}

\author{Alexei V. Filippenko}

\affil{Department of Astronomy, University of California, Berkeley, CA 
94720-3411 (e-mail: alex@astro.berkeley.edu)}

\begin{abstract}

   Nearly 4 years ago, two teams of observational astronomers reported that
high-redshift Type Ia supernovae are fainter than expected in a decelerating or
freely coasting universe.  The radical conclusion that the universe has been
accelerating in the past few billion years, possibly because of a nonzero value
for Einstein's cosmological constant, has gripped the worlds of astronomy and
physics, causing a flurry of new research. Having participated on both teams
(but much more closely with one than the other), here I provide a personal,
historical account of the story.

\end{abstract}


\section{INTRODUCTION}

  An accelerating universe! A nonzero value for Albert Einstein's cosmological
constant! Cosmic ``antigravity''! Not even a decade ago, who would have
seriously thought it? Certainly someone seeking tenure should not have made too
big a deal of this possibility!

   In retrospect, of course, some hints were already there (as summarized by
Ostriker \& Steinhardt 1995; see also Carroll, Press, \& Turner 1992). For
example, the calculated expansion age of the universe under the assumption that
the normalized mass density ($\Omega_M$) is equal to 1 [so that $t =
(2/3)H_0^{-1}$], even with a Hubble constant ($H_0$) as low as 65 km s$^{-1}$
Mpc$^{-1}$, is smaller than the derived ages of the oldest stars. Similarly,
numerical simulations of the growth of large-scale structure in the universe
are consistent with $\Omega \equiv \Omega_{\rm total} = 1$, but not if 70\% of
the mass-energy is in the form of hot dark matter. With the case for $\Omega_M
\approx 0.3$ becoming progressively stronger, a new culprit had to be found for
the remainder, if $\Omega$ really is equal to unity.

   But why should $\Omega = 1$, you might ask? Because the flatness problem
(i.e., we know that $\Omega$ is at least $\sim 0.1$ at the present time) is so
compelling, and inflation (or something similar in spirit, if not in detail)
provides a beautiful mechanism for the universe to achieve $\Omega =
1$. Although versions of inflation that don't demand $\Omega = 1$ were quickly
generated when the observational evidence for $\Omega_M \approx 0.3$ began to
pile up, the purists stuck to their guns: I remember a conference where Alan
Guth was confronted with the $\Omega_M \approx 0.3$ issue, and he responded by
saying that he didn't know what is going on, but he's confident that when all
the dust settles, observers will measure $\Omega = 1$. I do not, however,
recall that he explicitly mentioned the possibility of a nonzero cosmological
constant, $\Lambda$ --- a ``fudge factor'' that had reared its ugly head several
times in the past, only to be shot down by additional studies.  To many
physicists, the cosmological constant (like the anthropic principle) was the
last refuge of scoundrels.

   The initial, and most famous, appearance of $\Lambda$ came in 1917, when
Einstein tried to reconcile the basic property of gravity as we know it (which
pulls) with the apparently static universe.  But in the general theory of
relativity, what gravitates is the mass-energy density plus {\it 3 times the
pressure}, and if space has a sufficiently negative pressure (e.g., vacuum
energy, or some kind of rolling scalar field or ``quintessence''; e.g.,
Caldwell, Dav\'e, \& Steinhardt 1998), it will experience repulsion. A positive
cosmological constant, should it exist, would have the requisite negative
pressure; it is a vacuum energy whose density is constant, being a property of
space itself. Einstein set it equal to the precise value needed for a
stationary universe, even though the idea was repulsive (pun intended): there
was no other evidence for a nonzero value of $\Lambda$, such a value implied
that the spacetime curvature of completely empty space is nonzero, and it led
to a mathematically unstable solution (the universe had to be finely balanced
between expansion and contraction).

   In 1929, when Edwin Hubble announced that the current recession speeds of
galaxies are proportional to their distances, the entire physical and
philosophical motivation for a nonzero $\Lambda$ vanished; the universe is
expanding, not static. Einstein renounced his cosmological constant, calling it
the ``biggest blunder'' of his career; had he not insisted on its presence, he
could have predicted the dynamic nature of the universe before other physicists
such as Friedmann and Lema\^{i}tre had done so (though he might have expected the
universe to be contracting). But the cosmological constant itself is not a
mathematical blunder; rather, it is like a constant of integration whose value
must be determined. Einstein's ``blunder'' was in giving $\Lambda$ the precise
value needed for a static universe --- but he couldn't have known what would
transpire in the last decade of the millennium.

   The events that unfolded were dramatic indeed. Now, at the turn of the
millennium, the case for $\Omega_\Lambda \equiv \Lambda c^2/(3H_0^2) \approx
0.7$ is quite strong, though not yet certain. First, the data on high-redshift
Type Ia supernovae (SNe~Ia) were published (Riess et al. 1998b; Perlmutter et
al. 1999): high-redshift ($z \approx 0.5$) SNe~Ia are about 25\% fainter than
expected in a universe that has $\Omega_M = 0.3$, and $\sim$15\% fainter than
expected in a freely coasting universe ($\Omega_M = 0$), suggesting that the
expansion of the universe is accelerating with time. The data imply that
$\Omega_M - \Omega_\Lambda \approx -0.4$, so if $\Omega_M > 0.0$, then
$\Omega_\Lambda > 0.4$. Second, measurements of the predominant angular size of
fluctuations in the cosmic microwave background radiation (CMBR; de Bernardis
et al. 2000; Balbi et al. 2000; Hanany et al. 2000; Jaffe et al. 2001) show
that the spatial geometry of the universe is flat, and this requires $\Omega =
1.$ If $\Omega_M \approx 0.3$ (see, e.g., Bahcall 2000 for a summary), then
$\Omega_\Lambda \approx 0.7$. Third, the results of various deep galaxy
redshift surveys, most notably the Two-Degree Field Galaxy Redshift Survey
(Peacock et al. 2001; Efstathiou et al. 2001), are inconsistent with a universe
dominated by gravitating dark matter. The implication is that about 70\% of the
mass-energy density of the universe consists of some sort of vacuum energy or
``dark energy'' ({\it not} to be confused with dark matter!) whose
gravitational effect is repulsive --- a kind of ``cosmic antigravity.''
Wouldn't Einstein have been surprised to learn that his banished cosmological
constant had been resurrected --- not for the reason he had invoked (a static
universe), but with a different value, to account for an {\it accelerating}
universe!

   Here I present my personal view of the role that research on Type Ia
supernovae had on this exciting development. As the only person who at various
times was a member of both of the competing teams (the Supernova Cosmology
Project [SCP] and the High-$z$ Supernova Search Team [HZT]), I perhaps have a
unique perspective on the story. Of course, my view may not be free of biases,
especially given my much closer and more recent association with the HZT than
with the SCP. This is, by necessity, a brief summary; a much lengthier popular
account, as viewed by an outsider, has been published by Goldsmith (2000; see
also Livio 2000; Krauss 2000). There are numerous technical reviews, such as
those by Filippenko \& Riess (1998, 2000), Goldhaber \& Perlmutter (1998),
Goobar et al. (2000), Riess (2000), and Leibundgut (2001).

\section{EARLY HISTORY}

   Classical, hydrogen-deficient SNe~Ia are believed to result from the
explosions of white dwarfs in binary systems. In contrast, hydrogen-rich SNe~II
signal the deaths of massive stars through core collapse and neutrino-induced
rebound, perhaps in some cases aided by jets. (For a summary of observations,
see Filippenko 1997; recent theoretical reviews have been written by
Hillebrandt \& Niemeyer 2000 and Burrows \& Young 2000.] Both kinds of
explosions are very luminous, making them potentially visible at high
redshifts.

 The utility of SNe~Ia and SNe~II for determinations of the Hubble constant
is discussed by Branch (1998) and by Schmidt et al. (1994),
respectively.  But a measurement of the deceleration parameter ($q_0$) requires
SNe at much greater distances. Wagoner (1977) suggested that the distances of
SNe~Ia and SNe~II at redshift $z \approx 0.3$ could be measured with the
expanding photosphere method (Kirshner \& Kwan 1974), while Colgate (1979)
believed that the more luminous SNe~Ia, being nearly ``standard candles,''
could be used at $z \approx 1$ to determine the values of $q_0$ and
$\Lambda$. Tammann (1979) presented some of the fine points (e.g.,
K-corrections, host galaxy extinction, time dilation) that would need to be
considered for reliable results. Goobar \& Perlmutter (1995) explicitly showed
how $\Omega_M$ and $\Omega_\Lambda$ can be determined independently, given
enough SNe~Ia spanning a wide range of redshifts.

   The first long-term, serious attempt to use SNe~Ia for the determination of
$q_0$ was made by a Danish-led team using a 1.5-m telescope at the European
Southern Observatory (Hansen et al. 1989; N\o rgaard-Nielsen et al. 1989). They
obtained CCD images of clusters of galaxies repeatedly over the course of many
months, and searched for new objects (potential supernovae) in them by using
modern image-processing techniques. From their single definitive SN~Ia (SN
1988U, $z = 0.31$) they were able to constrain $q_0$ to be between $-0.6$ and
2.5, but the project was terminated after only about 2 years, largely because
of the low discovery rate. In a sense, the Danish group's project was attempted
too early (before adequate computing power was available), with a small
telescope and narrow field of view, and with early-generation CCDs.  But they
were important pioneers, showing in principle that their method of multi-epoch
imaging and careful data processing can lead to the discovery of high-redshift
SNe.

   In the early 1990s, Hamuy et al. (1993) used a similar technique of
repetitive imaging (in this case photographic) at the appropriate lunar phases
to conduct the very successful Cal\'an-Tololo search for relatively nearby
SNe. In addition, the Berkeley Automated Supernova Search Team, initiated by
Richard Muller and Carl Pennypacker of the Lawrence Berkeley National
Laboratory (LBNL), used a CCD camera on a small telescope at the University of
California, Berkeley's Leuschner Observatory to find many nearby supernovae
(Pennypacker et al.  1989; Perlmutter et al.  1992), among them the
well-observed SN 1990E (Schmidt et al. 1993). Studies of nearby SNe are very
important; cosmological parameters are deduced from {\it comparisons} of the
peak apparent brightnesses of SN at high redshifts and low redshifts.

\section{THE NEXT STEPS}

   Led by Saul Perlmutter of LBNL, the SCP formed in 1988, pushing forward the
original goal of Richard Muller and Carl Pennypacker to use high-redshift
SNe~Ia to measure $q_0$ (Pennypacker et al. 1989). This was a large
international team that included some astronomers, but its clear center of
activity was LBNL, and it consisted largely of people trained within the
physics community. I was not on the team when it first formed, but I joined in
1993 because of my astronomical expertise (especially in the field of SNe) and
my spectroscopic experience with large telescopes.  Specifically, I was to
conduct the spectroscopic confirmation and analysis of the SN~Ia candidates
with the Keck telescopes, and to offer advice on getting the best possible
results from the SNe.

   Although I was enthusiastic about participating in the collaboration, I did
not view it as the ``opportunity of a lifetime"; back then, SNe~Ia appeared to
be marginal standard candles, with a dispersion in peak luminosity ($\sim 0.5$
mag) about twice as large as the difference in apparent magnitude expected at
$z \approx 0.5$ in $q_0 = 0$ and $q_0 = 0.5$ universes.  Moreover, I thought
that at best, we would find $q_0 \approx 0.5$ ($\Omega_M \approx 1$), as
expected from the then-favorite theoretical model (the Einstein-deSitter
universe), but that the precision of our measurement would be insufficient to
determine whether the universe is open (and hence eternally expanding) or
closed (eventually collapsing).  This would especially be the case if standard
inflation dictated the early history of the universe: $\Omega_M$ would differ
from unity by an infinitesimal, unmeasurable amount. In addition, there were
differences in culture and perspective that I found difficult to overcome, but
I stayed with the team and contributed as best I could.

   To search for high-$z$ SNe, the SCP employed large-format CCDs on wide-angle
cameras attached to large-aperture telescopes; the first such imaging system
(with an $f/1$ focal reducer) was designed for the 3.9-m Anglo-Australian
Telescope. They also used newer, improved image-processing and analysis
techniques, having modified the software developed for their search for nearby
SNe (Pennypacker et al. 1989).  The SCP was the first group to show that SNe
could be found in ``batches'' (e.g., Perlmutter et al. 1994), at times
prescribed in advance so that sufficient follow-up observations could be
scheduled. This was a highly significant achievement. Initially it was
difficult for the team to secure sufficient telescope time for their search,
since they had not yet discovered any distant SNe --- yet to demonstrate
success, they needed access to large telescopes.  Moreover, there was growing
concern in some sectors of the supernova community that their project would not
yield reliable results even if high-$z$ SNe could be found; the measured
dispersion of nearby SNe~Ia when treated as perfect standard candles was $\sim
0.5$ mag (e.g., van den Bergh 1992; Branch \& Miller 1993), some extreme
deviants had been identified (Filippenko et al. 1992a,b; Phillips et al.  1992;
Leibundgut et al. 1993), and it was unclear whether the extinction would be
adequately taken into account.

  But the SCP pressed on, optimistic that with a growing understanding of
SNe~Ia, a method would be found to correct for the apparent heterogeneity of
SNe~Ia (other than to eliminate obvious outliers in the derived Hubble
diagram). They had crucial financial assistance and moral support from LBNL
(funded by the Department of Energy) and from the Center for Particle
Astrophysics at the University of California, Berkeley (an NSF Science and
Technology Center led by Bernard Sadoulet). Their first discovery was SN 1992bi
(Perlmutter et al. 1995a; $z = 0.458$ for the host galaxy; no spectrum of the
supernova itself was successfully obtained). This yielded a tentative
measurement of $q_0 = 0.1 \pm 0.3 \pm 0.55$ (assuming $\Lambda = 0$).

   An important paper by Mark Phillips (1993), who was a staff astronomer at
the Cerro-Tololo Interamerican Observatory (CTIO), led to a major improvement
in the cosmological utility of SNe~Ia. Using about 10 nearby, well-calibrated
SNe~Ia in galaxies of known distance, he showed that luminous SNe~Ia exhibit a
slower decline from maximum brightness than those having low peak
luminosity. Although a few previous astronomers had suggested such a relation,
it had always been viewed with suspicion because the data were poor. Phillips
used excellent light curves (derived in part by Nick Suntzeff, who was also at
CTIO) and showed beyond reasonable doubt that the relation exists. This paved
the way for precise distance measurements using SNe~Ia: the objects were not
exactly ``standard candles,'' but deviations from the nominal luminosity could
be taken into account by measuring the light-curve decline rate. The method was
further quantified by Riess, Press, \& Kirshner (1995), by Hamuy et
al. (1996a,b), and by Perlmutter et al. (1997). Moreover, by utilizing light
curves obtained through multiple filters, Riess, Press, \& Kirshner (1996)
showed that the SN extinction could be measured and removed in each individual
case.

  By 1994, the total number of high-$z$ SNe found by the SCP was 7, all
presumed to be SNe~Ia (although a few were not spectroscopically confirmed as
such).  The project began to gain considerable attention, both in the
astronomical community and the public eye (see, e.g., the PBS ``Mysteries of
Deep Space'' program, episode 2, ``Exploding Stars and Black Holes").  An
analysis of these first 7 objects suggested $\Omega_M = 0.88 \pm 0.6$ (for
$\Omega_\Lambda = 0$; or $\Omega_M = 0.94 \pm 0.3$ for $\Omega = 1$), a result
that was published by Perlmutter et al. (1997). Many astronomers,
however, were skeptical, especially those unfamiliar with the Phillips (1993)
relation and its subsequent refinements; they expressed significant
reservations about the cosmological utility of SNe~Ia.  Also, a high-density
universe seemed at odds with other, independent measurements of
$\Omega_M$. Indeed, although I was a member of the SCP, I was wary of our
conclusion; no corrections for extinction had been made, for example, and the
small sample size made it prone to errors produced by deviant SNe~Ia.

  Motivated in part by the Phillips (1993) relation for calibrating SNe~Ia, by
the scientific importance of a measurement of $q_0$, and by the success of the
SCP in finding high-$z$ SNe~Ia, a competing team (the HZT) was formed in 1994
by Brian P. Schmidt and Nick Suntzeff.  Schmidt had recently completed his
doctoral work under Bob Kirshner at the Harvard/Smithsonian Center for
Astrophysics (CfA), and has been at the Mt. Stromlo and Siding Spring
Observatories (Australia) since the beginning of 1995. Like the SCP, the HZT
was an international team --- but in contrast to the SCP, it consisted
primarily of astronomers, many of whom had made careers studying supernovae.
The HZT was structured in a less hierarchical manner than the SCP, which
followed the model often used by large-scale physics teams. The HZT had many
``generals'' (and, unfortunately, few ``soldiers'') loosely organized by the
young Schmidt, who was officially elected team leader in 1996.  Given what has
transpired in the past few years, it is clear that having two groups was very
beneficial to science --- progress was accelerated (pun intended) by the
competition, and results were more thoroughly checked for possible systematic
errors. If a potential bias was considered by one team and not the other, for
example, the second team would look bad.

  The HZT searched for high-$z$ SNe~Ia with the CTIO 4-m Blanco telescope ---
the equipment eventually adopted by the SCP after their hard-fought initial
success at the Anglo-Australian Telescope and the Isaac Newton Telescope. The
observing strategy was similar to that of the SCP, with some differences in
detail: obtain the first-epoch images just before first-quarter moon, the
second-epoch images just after third-quarter moon, and then commence follow-up
observations of identified SN candidates. Like the SCP (e.g., Perlmutter et
al. 1995b), the HZT demonstrated great success in finding batches of SNe,
sometimes over a dozen at a time (e.g., Suntzeff et al. 1996).  When possible,
the HZT's follow-up images were obtained through custom-made filters that
closely matched the $B$ and $V$ bands redshifted by 0.35 and 0.45, thereby
minimizing the K-corrections (Kim, Goobar, \& Perlmutter 1996).  Data reduction
and analysis were also done in a broadly similar fashion, with the HZT being
the first to stress proper accounting of reddening and extinction through the
use of the multi-color light-curve shape technique (MLCS; Riess et al. 1996).
The HZT found their first high-$z$ SN~Ia in 1995 (SN 1995K, $z = 0.48$; Schmidt
et al. 1998). Its spectrum is suggestive of a SN~Ia, though not completely
definitive. The light curves (Leibundgut et al. 1996), however, closely
resemble those of SNe~Ia, suitably dilated by a factor of $1+z$; in fact,
Wilson (1939) had proposed such a test for the expanding universe. During a
talk given in 1995, Goldhaber et al. (1997) also demonstrate time dilation in
the SCP light curves of SNe~Ia.

   In the Spring of 1996, I switched from the SCP to the HZT. Although I
continued to work with the SCP on some aspects of their project, such as the
reduction and analysis of Keck spectra of high-$z$ supernova candidates, my
primary allegiance was with the HZT. The switch occurred largely because of
differences in style and culture: I preferred to work within the somewhat
amorphous structure of the HZT, where my voice was more likely to be heard.
Also, the HZT's ways of resolving issues of scientific procedures and credit
were more to my liking. As was previously the case with the SCP, on the HZT I
was still largely responsible for the Keck spectroscopy of SN
candidates. However, I was also more closely involved with the cosmological
interpretation --- and indeed, a great opportunity presented itself when Adam
G. Riess, formerly Bob Kirshner's graduate student at the CfA, came to the
University of California, Berkeley in 1996 September as a Miller Postdoctoral
Fellow to work with me.

   One of Adam's first projects was to develop a quantitative method for
determining the age of a SN~Ia from its spectrum. His ``spectral feature age''
technique turned out to work remarkably well, and we were able to demonstrate
that the spectrum of SN 1996bj ($z = 0.57$) evolved more slowly by a factor of
$1 + z = 1.57$ than that of a nearby, low-redshift SN~Ia (Riess et
al. 1997). This effectively eliminated ``tired light'' and other non-expansion
hypotheses for the redshifts of objects at cosmological distances. (For
non-standard cosmological interpretations of all the SN~Ia data, see Narlikar
\& Arp (1997) and Hoyle, Burbidge, \& Narlikar 2000; a proper assessment of
these possible alternatives is beyond the scope of this essay.) Although one
might have been able to argue that something other than universal expansion
could be the cause of the apparent stretching of SN~Ia light curves at high
redshifts, it was much more difficult to attribute apparently slower evolution
of spectral details to an unknown effect. In a collaboration involving me,
Kirshner, and SCP members Perlmutter and Peter Nugent, Adam used spectral
feature ages to develop a method for determining ``snapshot distances'' of
SNe~Ia from just a single spectrum and a single night of multi-filter
photometry (Riess et al. 1998a). Such distances are slightly less precise than
those obtained from well-sampled SN light curves, but they have the advantage
of requiring much less telescope time.

\section{THE BREAKTHROUGH}

   In 1997, Adam was offered the opportunity to analyze and interpret all of
the HZT's data to date, including 16 high-$z$ SNe~Ia and the first object (SN
1995K; Schmidt et al.  1998). He had to work quickly and carefully, yet still
do a very thorough investigation. Competition from the SCP was stiff: they had
already published their $\Omega_M = 0.94 \pm 0.3$ (in a flat universe) result
based on 7 SNe~Ia (Perlmutter et al. 1997). Moreover, they set a redshift
record with SN 1997ap ($z = 0.83$), revising their estimate of $\Omega_M$ down
to $0.6 \pm 0.2$ (in a flat universe) and to $0.2 \pm 0.4$ if $\Omega_\Lambda =
0$ (Perlmutter et al. 1998), and they were busy analyzing their full set of 42
SNe~Ia. Meanwhile, HZT member Peter Garnavich, working as a postdoctoral fellow
with Kirshner at the CfA, was in charge of the analysis of three SNe~Ia for
which {\it Hubble Space Telescope (HST)} photometry was available. Among these
was SN 1997ck at $z = 0.97$, at that time a redshift record, although we cannot
be absolutely certain that the object was a SN~Ia because the spectrum is too
poor. From the three {\it HST} SNe~Ia and SN 1995K, Garnavich et al. (1998a)
found that $\Omega_M = 0.35 \pm 0.3$ (assuming $\Omega = 1$), or $\Omega_M =
-0.1 \pm 0.5$ (assuming $\Omega_\Lambda = 0$), inconsistent with the high
$\Omega_M$ initially found by Perlmutter et al. (1997) but consistent with the
revised estimate in Perlmutter et al. (1998).  However, none of these early
data sets carried the statistical discriminating power to detect cosmic
acceleration.

  Through the last few months of 1997, Adam's work on the HZT's full sample of
16 SNe~Ia progressed rapidly to completion.  During this time Adam and I often
discussed the HZT science, notably on the subjects of statistical errors,
potential systematic effects, and subtleties in the data calibration and
analysis.  In November 1997 the results seemed puzzling, indicating that if
$\Omega_\Lambda = 0$, $\Omega_M$ must be negative!  Adam initially checked his
work for simple errors (e.g., sign errors and programming bugs), not wanting to
reveal a silly error to the team.  By December 1997 it was clear that something
very strange had emerged from our data: the probable value of $\Omega_\Lambda$
was nonzero! My jaw just dropped when Adam showed me his Hubble diagram and
conclusions: the high-$z$ SNe~Ia were about 0.25 mag fainter than expected in a
low-density universe.  This was not the answer we had expected, and many
members of the HZT were worried that a subtle error had been made; indeed, Bob
Kirshner reflected that ``deep in our hearts, we know this can't be right.''
On the other side of the world, in Australia, HZT leader Brian Schmidt worked
hard to independently verify the analysis, making sure no obvious errors had crept
in.  A number of checks, however, did not reveal anything amiss --- if we were
wrong, it had to be for quite a subtle reason.  Moreover, we began to hear that
the SCP was also getting some odd, disturbing results!

   A press conference was scheduled at the 1998 January AAS meeting in
Washington, DC, with the stated purpose of presenting and discussing the
then-current evidence for a low-$\Omega_M$ universe as published by Perlmutter
et al. (1998; SCP) and Garnavich et al. (1998a; HZT). At the time, Adam did not
yet feel ready to announce the possible discovery of cosmic acceleration, since
various checks were still being made, and team member Peter Garnavich (who
represented the HZT at the press conference) was instructed not to mention
it. When showing the SCP's Hubble diagram for SNe~Ia, however, Saul Perlmutter
also pointed out tentative evidence for acceleration. He stated that the
conclusion was uncertain, and that the data were consistent with no
acceleration; consequently, members of the press generally did not emphasize
this result in their news reports. (James Glanz, in his article in the 1998
January 30 issue of {\it Science} magazine, was an exception.) But, of course,
members of the HZT did not fail to notice that the SCP's result pointed to the
same conclusion that Adam had made from the HZT data.

  In the next month, Adam worked hard to perform as many checks as possible of
the astonishing result. The conclusion that the universe is currently
accelerating, possibly because of a nonzero cosmological constant, did not go
away.  By February, Adam had completed a draft of the scientific paper
describing the results (Riess et al. 1998b).  Unable to find any significant
problem with the measurement, we decided that I would present the result at the
``Dark Matter '98'' meeting, to be held 1998 February 18--20 in Marina Del Rey,
California. Gerson Goldhaber and Saul Perlmutter spoke first, discussing the
SCP's demonstration of time dilation in the SN light curves, their strong
evidence for low $\Omega_M$, and their tentative evidence for nonzero
$\Omega_\Lambda$.  Then I gave a talk in which the HZT's results were shown,
and I stated that our data and extensive analysis strongly suggested that
$\Lambda$ is positive (Filippenko \& Riess 1998). There was a clear feeling of
excitement among the audience --- but also some disbelief and good, scientific
skepticism. Rocky Kolb of the University of Chicago, for example, mentioned
that the cosmological constant had come and gone at various other times in the
past century, and that the case here might be no different. He said there was
no obvious explanation for a value of $\Omega_\Lambda$ so small compared with
that expected from first principles ($10^{50} - 10^{120}$), and that a value of
precisely zero seemed much more likely. Later, when I gave a similar talk
elsewhere, a famous physics theorist told me that our observational results
{\it must} be wrong, since there was no conceivable way the cosmological
constant could differ infinitesimally from zero.

  Before the ``Dark Matter '98'' meeting, the HZT had not been planning to
issue a press release, and a paper had not yet been submitted to a refereed
journal. But rumors that we had found something very exciting had already been
leaked (not by members of the HZT) to at least one reporter.  I did not stick
around to talk to the press after my presentation at the ``Dark Matter '98''
meeting; instead, I flew to the Caribbean, as previously planned, to witness
the darkness of the totally eclipsed Sun. Upon my return, I found that in its
1998 February 27 issue, {\it Science} magazine had run a story by James Glanz
entitled ``Astronomers See a Cosmic Antigravity Force at Work.''  Brian Schmidt
was quoted as saying ``My own reaction is somewhere between amazement and
horror, amazement because I just did not expect this result, and horror in
knowing that [it] will likely be disbelieved by a majority of astronomers who,
like myself, are extremely skeptical of the unexpected.''

 Subsequently, other newspapers picked up on the story, and within a week it
had spread widely (e.g., {\it New York Times}, ``Wary Astronomers Ponder An
Accelerating Universe''). Adam Riess was busy fielding questions from the
press; he was even featured on the McNeil-Lehrer News Hour and CNN's Headline
News, and later in {\it TIME} magazine (2000 August) as one of the hottest
young astrophysicists to watch in the new millennium. By May of 1998, theorists
had organized a meeting in Chicago to discuss the startling results and the
nature of ``dark energy'' --- and the {\it New York Times} ran a story that
showed two of Michael Turner's viewgraphs (one titled ``Funny Energy in the
Univere'' [sic]).  Several new television documentaries featured the HZT and
SCP, including Equinox's ``Big G,'' the BBC Horizon's ``From Here to
Infinity,'' and, most recently and thoroughly, PBS's ``Runaway Universe''
(Nova).  At the Chicago meeting the results were debated, and in a straw poll
two-thirds of the attendees voted that they were convinced the results were
correct, in part because two independent teams had reached the same conclusion.

  The HZT's paper was officially accepted in May and published in the 1998
September issue of the {\it Astronomical Journal} (Riess et al. 1998b). With
the MLCS method applied to the full set of SNe~Ia, the formal results are
$\Omega_M = 0.24 \pm 0.10$ if $\Omega = 1$ (i.e., $\Omega_\Lambda = 0.76 \pm
0.10$, a $> 7\sigma$ detection), or $\Omega_M = -0.35 \pm 0.18$ (which is
unphysical) if $\Omega_\Lambda = 0$. The confidence contours in the $\Omega_M$
vs. $\Omega_\Lambda$ plane (Riess et al. 1998b) suggest that $\Omega_\Lambda >
0$ at the $\sim 3\sigma$ level; the precise results depend on the method used
to analyze the light curves of SNe (MLCS, or ``$\Delta m_{\rm 15}$'' which is
based on the total decline in the first 15 days past maximum brightness), but
they are consistent with each other. The dynamical age of the universe could
then be calculated from the cosmological parameters; the result is about 14.2
Gyr if $H_0 = 65$ km s$^{-1}$ Mpc$^{-1}$ (Riess et al.  1998b). This age is
consistent with values determined from various other techniques; specifically,
the recently revised ages of globular star clusters (about 13 Gyr) were no
longer bigger than the expansion age of the universe. The HZT also concluded
that the SN~Ia data, when combined with the then-available measurements of the
CMBR, show that $\Omega = 0.94 \pm 0.26$ (Garnavich et al. 1998b), consistent
with a flat universe.

   From an essentially independent set of 42 high-$z$ SNe~Ia (only 2 objects in
common), the SCP later published their almost identical conclusions (Perlmutter
et al. 1999). (Although the SCP included more SNe in their study, the error
bars per object were larger than in the HZT's work, so the final answer had
roughly comparable uncertainty.) This agreement suggests that neither team had
made a large, simple blunder! If the result was wrong, the reason had to be
subtle.

   In its 1998 December 18 issue, {\it Science} magazine named the HZT's and
SCP's co-discovery of an accelerating universe the top ``Science Breakthrough
of 1998.''  Although both teams were honored to have been recognized in this
manner, they were not yet certain that they were right. However, the editors of
{\it Science} magazine noted that rarely is an important discovery made and
confirmed beyond reasonable doubt within the same year. About 3/4 of a year had
passed since the announcement of acceleration, and nobody had shot definitive
holes in the analysis, so the editors of {\it Science} magazine felt justified
in their proclamation. Their magazine cover showed a caricature of Einstein
blowing a ``bubble universe'' out of his pipe and watching (with a very
surprised expression) its expansion accelerate, while holding a sheaf of papers
on which $\Lambda$ can be seen. Einstein had renounced the cosmological
constant in 1929, following Edwin Hubble's discovery of the expansion of the
universe, but now the HZT and SCP had resurrected it --- to produce not a
static universe, but rather an accelerating one. Yes, he would surely have been
quite surprised to learn of this development, were he alive now!

\section{SEARCHING FOR SYSTEMATIC ERRORS}

  The conclusion reached by the HZT and SCP is so dramatic, so ``crazy'' in
some respects, that it behooves us to find an alternative explanation --- a
systematic effect that causes high-$z$ SNe~Ia to appear fainter than
expected. Although both teams had considered a number of potential systematic
effects in their discovery papers (Riess et al. 1998b; Perlmutter et al. 1999),
and had shown with reasonable confidence that obvious ones were not greatly
affecting their conclusions, it was of course possible that they were wrong,
and that some other culprit was leading to an incorrect interpretation of the
data. The two obvious effects are cosmic evolution of the peak luminosity of
SNe~Ia (i.e., they were intrinsically dimmer in the past), and relatively gray
extinction (since normal extinction had been taken into account by using the
MLCS technique).

  One way to test for cosmic evolution of SNe~Ia is to compare all measurable
properties of low-$z$ and high-$z$ SNe~Ia, and see if they differ. If they
don't, then a reasonable (but not absolutely iron-clad) conclusion is that
their peak luminosities are also the same. For example, the HZT showed that the
spectrum of a particularly well-observed SN~Ia at $z = 0.45$ is very similar to
that of a nearby SN~Ia (Coil et al. 2000; see also Perlmutter et al. 1998;
Riess et al. 1998b). Moreover, Riess et al. (2000) showed that the restframe
near-infrared light curve of SN 1999Q ($z = 0.46$) probably has a second
maximum about a month after the first one, just like that of nearby SNe~Ia of
normal luminosity, and unlike subluminous SNe~Ia such as SN 1991bg (Filippenko
et al. 1992b). Additional tests with spectra and near-infrared light curves are
currently being conducted.

  Another way of using light curves to test for possible evolution of SNe~Ia is
to see whether the risetime (from explosion to maximum brightness) is the same
for high-$z$ and low-$z$ SNe~Ia; a difference might indicate that the peak
luminosities are also different. Though the exact value of the risetime is a
function of peak luminosity, for typical low-$z$ SNe~Ia it is $20.0 \pm 0.2$
days (Riess et al. 1999b). We pointed out (Riess et al. 1999a) that this
differs by $5.8\sigma$ from the {\it preliminary} risetime of $17.5 \pm 0.4$
days previously reported in conferences by the SCP (Goldhaber et al. 1998;
Groom 1998).  However, a more thorough analysis of the SCP data
(Aldering, Knop, \& Nugent 2000) shows that the high-$z$ uncertainty of $\pm
0.4$ days that the SCP originally reported was too small because it did
not account for unappreciated systematic effects and correlated errors. The
revised discrepancy with the low-$z$ risetime is about $2\sigma$ or less. Thus,
the apparent difference in risetimes might be insignificant. Even if the
difference is real, however, its relevance to the peak luminosity is unclear;
the light curves may differ only in the first few days after the explosion, and
this could be caused by small variations in conditions near the outer part of
the exploding white dwarf that are inconsequential at the peak.

   Let us now consider the possibility of extinction. As mentioned above, our
procedure already corrects for extinction produced by normal dust grains
similar to the average grains in the Galaxy. However, could an evolution in
dust grain size descending from ancestral interstellar ``pebbles'' at higher
redshifts cause us to underestimate the extinction (e.g, Aguirre 1999a,b)?
Large dust grains would not imprint the reddening signature of typical
interstellar extinction upon which our corrections rely. But even the dust
postulated by Aguirre is not completely gray, having a minimum size of
about 0.1~$\mu$m. We can test for such nearly gray dust by observing
high-redshift SNe~Ia over a wide wavelength range to measure the color excess
it would introduce. If $A_V = 0.25$ mag, then $E(U-I)$ and $E(B-I)$ should be
0.12--0.16 mag. If, on the other hand, the 0.25 mag faintness is due to
$\Lambda$, then no such reddening should be seen.  This effect is measurable
using proven techniques; so far, with just one SN~Ia (SN 1999Q, $z = 0.46$),
the HZT's results favor the no-dust hypothesis to better than 2$\sigma$ (Riess
et al. 2000).  More work along these lines is in progress, but not without
obstacles: a 5-night HZT observing run at Keck in Fall 2000, for spectroscopic
identification of high-$z$ SN candidates, was completely washed out by bad
weather! HZT team members Bruno Leibundgut and Jesper Sollerman came to the
rescue with two clear nights at ESO's Very Large Telescope, the second of which
had $\sim 0.3''$ seeing.

\section{HINTS OF THE SMOKING GUN}

   The most decisive test to distinguish between $\Lambda$ and cumulative
systematic effects, however, is to examine the {\it deviation} of the observed
peak magnitude of SNe~Ia from the magnitude expected in the low-$\Omega_M$,
zero-$\Lambda$ model. If $\Lambda$ is positive, the deviation should actually
begin to {\it decrease} at $z \approx 1$; we will be looking so far back in
time that the $\Lambda$ effect becomes small compared with $\Omega_M$, and the
universe is decelerating at that epoch.  If, on the other hand, a systematic
bias such as gray dust or evolution of the white dwarf progenitors is the
culprit, we expect that the deviation of the apparent magnitude will continue
growing (for example, see Figure 11 in Filippenko \& Riess 2000, or Figure 13
in Riess 2000), unless the systematic bias is set up in such an unlikely way as
to mimic the effects of $\Lambda$ (e.g., Drell, Loredo, \& Wasserman 2000).  A
turnover, or decrease of the deviation of apparent magnitude at high redshift,
can be considered the ``smoking gun'' of $\Lambda$.  Thus, the HZT and SCP have
embarked on campaigns to find and monitor SNe~Ia at $z \gtrsim 0.8$. Results
for two SNe~Ia at $z > 1$ measured by the HZT already look promising (J. Tonry
et al., in preparation); their deviation in apparent magnitude is roughly the
same as that at $z \approx 0.5$. The data would have been even more convincing
had {\it HST} not lost its third gyro in Fall 1999, placing it in ``safe mode''
(and hence unusable!) during a critical time in our program.

  Very recently, Riess et al. (2001) reported {\it HST} observations of a
probable SN~Ia at $z \approx 1.7$ (the most distant SN~Ia ever observed) that
suggest the expected turnover is indeed present, providing a tantalizing
glimpse of the epoch of deceleration (Riess et al. 2001). This object, SN
1997ff, was discovered by Gilliland \& Phillips (1998) in a repeat {\it HST}
observation of the Hubble Deep Field--North, and serendipitously monitored in
the infrared with {\it HST}/NICMOS. The peak apparent SN brightness is
consistent with that expected in the decelerating phase of the preferred
cosmological model, $\Omega_M \approx 0.3, \Omega_\Lambda \approx 0.7$. It is
inconsistent with gray dust or simple luminosity evolution, candidate
astrophysical effects which could mimic previous evidence for an accelerating
universe from SNe~Ia at $z \approx 0.5$.

   The possible discovery of a turnover, as well as complementary studies such
as those of the CMBR, led to the cover story on the 2001 June 25 issue of {\it
TIME} magazine: ``How the Universe will End.'' (Adam Riess keeps good company
in the story, being singled out along with Hubble, Einstein, Zwicky, Penzias,
and Wilson as giants who studied the universe.) On the other hand, it is wise
to remain cautious: the error bars are large, and it is always possible that we
are being fooled by this one object.  Clearly, more SNe~Ia at such high
redshifts should be found and monitored in the future to help verify the
hypothesis of a currently accelerating and previously decelerating universe.

  Another very important question to address is whether the ``dark energy" is
caused by a cosmological constant or some other phenomenon such as quintessence
(e.g., Caldwell et al. 1998).  If $\Lambda$ dominates, then the equation of
state of the dark energy should have an index $w = -1$, where the pressure
($P$) and density ($\rho$) are related according to $w = P/(\rho c^2)$.
Garnavich et al. (1998b) and Perlmutter et al. (1999) already set an
interesting limit, $w \lesssim -0.60$ at the 95\% confidence level. However,
more high-quality data at $z \approx 0.5$ are needed to narrow the allowed
range.

   Farther in the future, large numbers of SNe~Ia found by the {\it
Supernova/Acceleration Probe} ({\it SNAP}; Nugent 2000) and the Large-area
Synoptic Survey Telescope (the ``Dark Matter Telescope''; Tyson \& Angel 2001)
could reveal whether the value of $w$ depends on redshift, and hence should
give additional constraints on the nature of the dark energy. High-redshift
surveys of galaxies, such as DEEP2 (Davis et al. 2001), should provide
independent evidence for (or against!) $\Lambda$. And, of course, the
space-based missions to map the CMBR (e.g., MAP, Planck) are designed to obtain
a wealth of valuable data.  We've already come a long way, but these projects
and others promise to provide much future excitement as well. I never thought
that I would be involved in such a fundamental development during my career,
and I am eternally grateful for the opportunity to have contributed. It's been
a blast.

\acknowledgments

   The following grants have generously supported my own research with the SCP and
especially the HZT: NSF AST-9417213 and AST-9987438, as well as NASA/{\it HST}
GO-7505, GO-7588, GO-8177, GO-8641, and GO-9118 from the Space Telescope
Science Institute, which is operated by the Association of Universities for
Research in Astronomy, Inc., under NASA contract NAS~5-26555.  I also
acknowledge the Guggenheim Foundation for a Fellowship. The results summarized
here would not have been possible without the hard work of members of the HZT
and SCP, and of many other astronomers. I am especially grateful to Adam Riess
for countless stimulating discussions, to Brian Schmidt for excellent
leadership of the HZT, and to Saul Perlmutter for pioneering the ``batch" 
discovery of high-redshift supernovae.

\end{document}